\documentclass[twocolumn,amsmath,amssymb]{revtex4}
\usepackage{graphicx}
\usepackage{subfigure}
\usepackage{dcolumn}
\usepackage{bm}
\usepackage{color}
\usepackage{pstricks}
\usepackage{pst-text}
\begin{document}
\title{Upper Bound for Critical Probability of Site Percolation on Triangular Lattice is \(1/2^{\frac{3}{2}}\)}
\author{Marko Puljic}
\email[Marko Puljic: ]{mpuljic@tulane.edu}
\affiliation{Center for Computational Science at Tulane University}
\date{\today}
\date{\today}
\begin{abstract}
In triangular lattice, the upper bound for critical probability of site percolation is \(\displaystyle\frac{1}{2^{\frac{3}{2}}}\approx0.3535\).
\end{abstract}

\maketitle
\maketitle

\section{Introduction and Description of \(\mathbb{Z}^{2}\)}
\begin{figure}
\definecolor{gry}{gray}{.9}
\begin{pspicture}(4,4.2)(0,0)

\psline[linewidth=6pt,linestyle=solid,linecolor=gry]{-}(-0.2,4)(4.2,4)
\psline[linewidth=6pt,linestyle=solid,linecolor=gry]{-}(0.3,3.5)(3.7,3.5)
\psline[linewidth=6pt,linestyle=solid,linecolor=gry]{-}(0.8,3)(3.2,3)
\psline[linewidth=6pt,linestyle=solid,linecolor=gry]{-}(1.3,2.5)(2.7,2.5)

\psline[linewidth=0.5pt,linestyle=solid]{-}(0,3)(1,4)
\psline[linewidth=0.5pt,linestyle=solid]{-}(0,2)(2,4)
\psline[linewidth=0.5pt,linestyle=solid]{-}(0,1)(3,4)
\psline[linewidth=0.5pt,linestyle=solid]{-}(0,0)(4,4)
\psline[linewidth=0.5pt,linestyle=solid]{-}(1,0)(4,3)
\psline[linewidth=0.5pt,linestyle=solid]{-}(2,0)(4,2)
\psline[linewidth=0.5pt,linestyle=solid]{-}(3,0)(4,1)

\psline[linewidth=0.5pt,linestyle=solid]{-}(0,1)(1,0)
\psline[linewidth=0.5pt,linestyle=solid]{-}(0,2)(2,0)
\psline[linewidth=0.5pt,linestyle=solid]{-}(0,3)(3,0)
\psline[linewidth=0.5pt,linestyle=solid]{-}(0,4)(4,0)
\psline[linewidth=0.5pt,linestyle=solid]{-}(1,4)(4,1)
\psline[linewidth=0.5pt,linestyle=solid]{-}(2,4)(4,2)
\psline[linewidth=0.5pt,linestyle=solid]{-}(3,4)(4,3)

{\color{white}
\multiput(0,0)(2,0){2}{\circle*{0.15}} \put(3,0){\circle*{0.15}}
\put(2.5,0.5){\circle*{0.15}}
\multiput(1,1)(2,0){2}{\circle*{0.15}}
\multiput(1.5,1.5)(2,0){2}{\circle*{0.15}}
\multiput(0,2)(1,0){2}{\circle*{0.15}}
\multiput(2.5,2.5)(1,0){2}{\circle*{0.15}}
\multiput(1,3)(3,0){2}{\circle*{0.15}}
\put(2.5,3.5){\circle*{0.15}}
\multiput(0,4)(2,0){3}{\circle*{0.15}}
}
\multiput(0,0)(2,0){2}{\circle{0.15}} \put(3,0){\circle{0.15}}
\put(2.5,0.5){\circle{0.15}}
\multiput(1,1)(2,0){2}{\circle{0.15}}
\multiput(1.5,1.5)(2,0){2}{\circle{0.15}}
\multiput(0,2)(1,0){2}{\circle{0.15}}
\multiput(2.5,2.5)(1,0){2}{\circle{0.15}}
\multiput(1,3)(3,0){2}{\circle{0.15}}
\put(2.5,3.5){\circle{0.15}}
\multiput(0,4)(2,0){3}{\circle{0.15}}

\multiput(1,0)(3,0){2}{\circle*{0.15}}
\multiput(0.5,0.5)(1,0){2}{\circle*{0.15}} \put(3.5,0.5){\circle*{0.15}}
\multiput(0,1)(2,0){3}{\circle*{0.15}}
\multiput(0.5,1.5)(2,0){2}{\circle*{0.15}}
\multiput(3,2)(1,0){2}{\circle*{0.15}}
\multiput(0.5,2.5)(1,0){2}{\circle*{0.15}}
\multiput(0,3)(3,0){2}{\circle*{0.15}} \put(2,3){\circle*{0.15}}
\multiput(0.5,3.5)(1,0){2}{\circle*{0.15}} \put(3.5,3.5){\circle*{0.15}}
\multiput(1,4)(2,0){2}{\circle*{0.15}}

\put(2,2){\circle*{0.2}}

\psline[linewidth=0.5pt,linestyle=dotted,linearc=0.45]{-}(0.2,3.8)(3.8,3.8)(3.8,0.2)(3.8,0.2)(0.2,0.2)(0.2,3.8)

{\footnotesize
\put(1.93,2.17){0}
\put(2.25,2.8){\(\mathbf 1\!\!\uparrow_{1}\)}
\put(2.75,3.3){\(\mathbf 2\!\!\uparrow_{1}\)}
\put(3.25,3.8){\(\mathbf 3\!\!\uparrow_{1}\)}
\put(2.25,1.8){\(\mathbf 1\!\!\downarrow_{2}\)}
\put(2.75,1.3){\(\mathbf 2\!\!\downarrow_{2}\)}
\put(3.25,0.8){\(\mathbf 3\!\!\downarrow_{2}\)}
\put(1.25,2.8){\(\mathbf 1\!\!\uparrow_{2}\)}
\put(0.75,3.3){\(\mathbf 2\!\!\uparrow_{2}\)}
\put(0.25,3.8){\(\mathbf 3\!\!\uparrow_{2}\)}
\put(1.25,1.8){\(\mathbf 1\!\!\downarrow_{1}\)}
\put(0.75,1.3){\(\mathbf 2\!\!\downarrow_{1}\)}
\put(0.25,0.8){\(\mathbf 3\!\!\downarrow_{1}\)}

\put(1.45,4.12){\(\mathbf 2\!\uparrow_{1}\!\!+\mathbf 2\!\uparrow_{2}\)}
}

\end{pspicture}
\caption{\label{lattice2Deg5}
In arcs, \(\sum_{i}|a_{i}(\mathbf v)|=4\) (outside dotted curve).
In the convenient pair of opposite arcs \(\{\mathcal A_{4}(\mathbb{Z}^{2}),\mathcal A_{-4}(\mathbb{Z}^{2})\}_{1}\), \(\mathcal A_{4}(\mathbb{Z}^{2})=\{4\!\uparrow_{1},3\!\uparrow_{1}\!\!+1\!\uparrow_{2},2\!\uparrow_{1}\!\!+2\!\uparrow_{2},1\!\uparrow_{1}\!\!+3\!\uparrow_{2},4\!\uparrow_{2}\}\).
}
\end{figure}
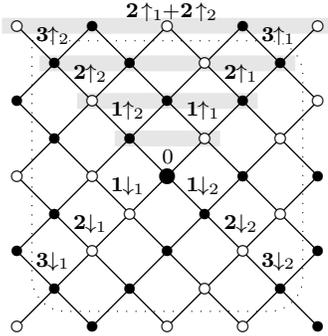

\noindent In site percolation, vertices (sites) of a graph are open with probability \(p\), and there is the smallest \(p=p_{H}\), critical \(p\), for which open vertices form an open path the long way across a graph, so a vertex at the origin is a part of an infinite connected open vertex set, \footnote{Path is a walk via edges visiting each vertex only once.}.
Smirnov found that for triangular lattice \(p_{H}=1/2\), \citep{smirnov2001}, but there is the traversal, from the origin upwards, so that an infinite connected open vertex set exists for \(p_{H}=1/2^{\frac{3}{2}}\approx0.3535\).
\\ \\
In finite graph \(\mathbb{Z}_{k}^{2}\), \(2\) pairs of opposite arcs are \(k\) edges away from the vertex at origin, Fig. \ref{lattice2Deg5}.
Basis \(\mathcal B(\mathbb{Z}^{2})\) of edges and the integers \(a_{i}\) assign the place to vertex \(\mathbf v\in\mathbb{Z}_{k}^{2}\):
\begin{align}
\mathbf v_{0}&=0=\text{vertex at origin of }\mathbb{Z}_{k}^{2}\subset\mathbb{Z}^{2} \notag \\
\mathcal B\left(\mathbb{Z}^{d}\right)&=\left\{\uparrow_{1},\uparrow_{2},\uparrow_{3},..,\uparrow_{d}\right\},\ -\!\uparrow_{1}=\downarrow_{1}\ \&\ -\!\uparrow_{2}=\downarrow_{2} \notag \\
\mathbf v&=a_{1}\!\uparrow_{1}+\ a_{2}\!\uparrow_{2},\ \mathbf v_{0}\to\mathbf v=\text{path from \(\mathbf v_{0}\) to \(\mathbf v\)} \notag \\
|\mathbf v_{0}\to\mathbf v|&=\lVert\mathbf v\rVert=|a_{1}(\mathbf v)|+|a_{2}(\mathbf v)| \notag
\end{align}
Places of neighbors of vertex \(\mathbf v\in\mathbb{Z}^{2}\) can be partition into \(2\) up-step neighbors traversed via \(\uparrow_{1}\) and \(\uparrow_{2}\) and \(2\) down-step neighbors traversed via \(\downarrow_{1}\) and \(\downarrow_{2}\).
For \(\uparrow_{1}\), there is matching \(\downarrow_{1}\), and for \(\uparrow_{2}\), there is matching \(\downarrow_{2}\):
\begin{align}
\mathcal N(\mathbf v,\mathbb{Z}^{2})&=\text{neighbors of }\mathbf v \notag \\
\mathcal N_{u}(\mathbf v,\mathbb{Z}^{2})&=\text{up-step neighbors of }\mathbf v=\{\mathbf v+\!\!\uparrow_{1},\mathbf v+\!\!\uparrow_{2}\} \notag
\end{align}
\(\mathbf v_{0}\) is \(k\) edges away from the arcs and it can be any vertex in \(\mathbb{Z}^{2}\).
Two pairs of opposite arcs in \(\mathbb{Z}_{k}^{2}\) look the same and any pair, by rotation of \(\mathbb{Z}^{2}\), can be a convenient pair \(\{\mathcal A_{k}(\mathbb{Z}^{2}),\mathcal A_{-k}(\mathbb{Z}^{2})\}_{1}\):
\begin{align}
\left\{\mathcal A_{k}(\mathbb{Z}^{2}),\mathcal A_{-k}(\mathbb{Z}^{2})\right\}_{1}&=\text{one of }2\text{ pairs of opposite arcs} \notag \\
\mathcal A_{k}(\mathbb{Z}^{2})\cap\mathcal A_{-k}(\mathbb{Z}^{2})&=\emptyset \notag
\end{align}
\begin{align}
\mathcal A_{k}(\mathbb{Z}^{2})&=\left\{
\begin{array}{c}
\mathbf v: \lVert\mathbf v\rVert=k\ \&\ a_{1}(\mathbf v),a_{2}(\mathbf v)\ge0 \\
\text{or} \\
\mathbf v\text{ built with }\uparrow_{1}\ \&\ \uparrow_{2}\text{ edges only}
\end{array}
\right\}
\notag
\end{align}
The shortest traversal from \(\mathbf v_{0}\) to \(\mathcal A_{k}(\mathbb{Z}^{2})\) is a traversal via vertices in \(\mathcal N_{u}(\mathbf v,\mathbb{Z}^{2})\), which are one edge closer to \(\mathcal A_{k}(\mathbb{Z}^{2})\):
\begin{align}
1^{st}&\!:\mathcal A_{1}\left(\mathbb{Z}^{2}\right)=\{\uparrow_{1},\uparrow_{2}\} \notag \\
2^{nd}&\!:\mathcal A_{2}\left(\mathbb{Z}^{2}\right)=\!\!\!\!\displaystyle\bigcup_{\mathbf v\in\mathcal A_{1}(\mathbb{Z}^{2})}\!\!\!\!\mathcal N_{u}(\mathbf v,\mathbb{Z}^{2}) \notag \\
.. \notag \\
k^{th}&\!:\mathcal A_{k}\left(\mathbb{Z}^{2}\right)=\!\!\!\!\displaystyle\bigcup_{\mathbf v\in\mathcal A_{k-1}(\mathbb{Z}^{2})}\!\!\!\!\mathcal N_{u}(\mathbf v,\mathbb{Z}^{2}) \notag
\end{align}

\section{Triangular Lattice Percolation}
\noindent Lattice \(\mathbb{Z}^{2}\) is percolating when the open vertices form an open path connecting \(\mathbf v_{0}\) with the vertices in the opposite sides of \(\mathbb{Z}_{k}^{2}\) and \(k\to\infty\), so we need to know the number of paths connecting \(\mathbf v_{0}\) and vertices in \(\mathcal A_{k}(\mathbb{Z}^{2})\):
\begin{align}
p&=\text{probability of vertex being open} \notag \\
\psi\left(\mathbb{Z}^{2},p\right)&=\text{number of percolating paths in \(\mathbb{Z}^{2}\)} \notag
\end{align}
\\ \\
\noindent 
\(\mathbb{Z}^{2}\) is embedded in \(\mathbf T\), which has two pairs of opposite sides and two definitions of \(\mathcal B(\mathbf T)\), Fig. \ref{latticeTexample}:
\begin{align}
\mathbf T&=\text{triangular lattice with embedded }\mathbb{Z}^{2} \notag \\
\mathcal B(\mathbf T)&=
\left\{
\begin{array}{l}
\uparrow_{1},\uparrow_{2},\uparrow_{1}\!\!+\!\!\uparrow_{2}=\uparrow_{1,2}=\!-\!\downarrow_{1,2} \\
\ \ \ \ \ \ \text{ or} \\
\uparrow_{1},\uparrow_{2},\uparrow_{1}\!\!+\!\!\downarrow_{2}=\rightarrow_{1,2}=\!-\!\leftarrow_{1,2}
\end{array}
\right\}
\notag \\
&\lVert\uparrow_{1,2}\rVert=\lVert\rightarrow_{1,2}\rVert=1 \notag \\
\mathcal N_{u}(\mathbf v,\mathbf T)&=
\left\{
\begin{array}{l}
\mathbf v+\!\!\uparrow_{1},\mathbf v+\!\!\uparrow_{2},\mathbf v+\!\!\uparrow_{1,2} \\
\ \ \ \ \ \ \text{ or} \\
\mathbf v+\!\!\uparrow_{1},\mathbf v+\!\!\uparrow_{2} \\
\end{array}
\right\}
\notag
\end{align}
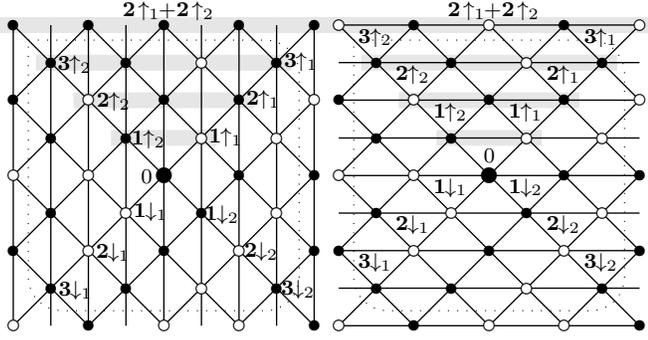
\begin{figure}
\definecolor{gry}{gray}{.9}
\begin{pspicture}(4,4.2)(0,0)

\psline[linewidth=6pt,linestyle=solid,linecolor=gry]{-}(-0.2,4)(4.2,4)
\psline[linewidth=6pt,linestyle=solid,linecolor=gry]{-}(0.3,3.5)(3.7,3.5)
\psline[linewidth=6pt,linestyle=solid,linecolor=gry]{-}(0.8,3)(3.2,3)
\psline[linewidth=6pt,linestyle=solid,linecolor=gry]{-}(1.3,2.5)(2.7,2.5)

\psline[linewidth=0.5pt,linestyle=solid]{-}(0,3)(1,4)
\psline[linewidth=0.5pt,linestyle=solid]{-}(0,2)(2,4)
\psline[linewidth=0.5pt,linestyle=solid]{-}(0,1)(3,4)
\psline[linewidth=0.5pt,linestyle=solid]{-}(0,0)(4,4)
\psline[linewidth=0.5pt,linestyle=solid]{-}(1,0)(4,3)
\psline[linewidth=0.5pt,linestyle=solid]{-}(2,0)(4,2)
\psline[linewidth=0.5pt,linestyle=solid]{-}(3,0)(4,1)

\psline[linewidth=0.5pt,linestyle=solid]{-}(0,1)(1,0)
\psline[linewidth=0.5pt,linestyle=solid]{-}(0,2)(2,0)
\psline[linewidth=0.5pt,linestyle=solid]{-}(0,3)(3,0)
\psline[linewidth=0.5pt,linestyle=solid]{-}(0,4)(4,0)
\psline[linewidth=0.5pt,linestyle=solid]{-}(1,4)(4,1)
\psline[linewidth=0.5pt,linestyle=solid]{-}(2,4)(4,2)
\psline[linewidth=0.5pt,linestyle=solid]{-}(3,4)(4,3)

\multiput(0,0)(0.5,0){9}{\psline[linewidth=0.5pt,linestyle=solid]{-}(0,0)(0,4)}

{\color{white}
\multiput(0,0)(2,0){2}{\circle*{0.15}} \put(3,0){\circle*{0.15}}
\put(2.5,0.5){\circle*{0.15}}
\multiput(1,1)(2,0){2}{\circle*{0.15}}
\multiput(1.5,1.5)(2,0){2}{\circle*{0.15}}
\multiput(0,2)(1,0){2}{\circle*{0.15}}
\multiput(2.5,2.5)(1,0){2}{\circle*{0.15}}
\multiput(1,3)(3,0){2}{\circle*{0.15}}
\put(2.5,3.5){\circle*{0.15}}
}
\multiput(0,0)(2,0){2}{\circle{0.15}} \put(3,0){\circle{0.15}}
\put(2.5,0.5){\circle{0.15}}
\multiput(1,1)(2,0){2}{\circle{0.15}}
\multiput(1.5,1.5)(2,0){2}{\circle{0.15}}
\multiput(0,2)(1,0){2}{\circle{0.15}}
\multiput(2.5,2.5)(1,0){2}{\circle{0.15}}
\multiput(1,3)(3,0){2}{\circle{0.15}}
\put(2.5,3.5){\circle{0.15}}

\multiput(1,0)(3,0){2}{\circle*{0.15}}
\multiput(0.5,0.5)(1,0){2}{\circle*{0.15}} \put(3.5,0.5){\circle*{0.15}}
\multiput(0,1)(2,0){3}{\circle*{0.15}}
\multiput(0.5,1.5)(2,0){2}{\circle*{0.15}}
\multiput(3,2)(1,0){2}{\circle*{0.15}}
\multiput(0.5,2.5)(1,0){2}{\circle*{0.15}}
\multiput(0,3)(3,0){2}{\circle*{0.15}} \put(2,3){\circle*{0.15}}
\multiput(0.5,3.5)(1,0){2}{\circle*{0.15}} \put(3.5,3.5){\circle*{0.15}}
\multiput(0,4)(1,0){5}{\circle*{0.15}}

\put(2,2){\circle*{0.2}}

\psline[linewidth=0.5pt,linestyle=dotted,linearc=0.45]{-}(0.2,3.8)(3.8,3.8)(3.8,0.2)(3.8,0.2)(0.2,0.2)(0.2,3.8)

{\footnotesize
\put(1.7,1.9){0}
\put(2.6,2.44){\(\mathbf 1\!\!\uparrow_{1}\)}
\put(3.1,2.94){\(\mathbf 2\!\!\uparrow_{1}\)}
\put(3.6,3.44){\(\mathbf 3\!\!\uparrow_{1}\)}
\put(2.55,1.41){\(\mathbf 1\!\!\downarrow_{2}\)}
\put(3.07,0.92){\(\mathbf 2\!\!\downarrow_{2}\)}
\put(3.56,0.42){\(\mathbf 3\!\!\downarrow_{2}\)}
\put(1.57,2.41){\(\mathbf 1\!\!\uparrow_{2}\)}
\put(1.11,2.91){\(\mathbf 2\!\!\uparrow_{2}\)}
\put(0.59,3.41){\(\mathbf 3\!\!\uparrow_{2}\)}
\put(1.6,1.45){\(\mathbf 1\!\!\downarrow_{1}\)}
\put(1.1,0.9){\(\mathbf 2\!\!\downarrow_{1}\)}
\put(0.6,0.4){\(\mathbf 3\!\!\downarrow_{1}\)}

\put(1.45,4.12){\(\mathbf 2\!\uparrow_{1}\!\!+\mathbf 2\!\uparrow_{2}\)}
}

\end{pspicture}
\definecolor{gry}{gray}{.9}
\begin{pspicture}(4,4.2)(0,0)

\psline[linewidth=6pt,linestyle=solid,linecolor=gry]{-}(-0.2,4)(4.2,4)
\psline[linewidth=6pt,linestyle=solid,linecolor=gry]{-}(0.3,3.5)(3.7,3.5)
\psline[linewidth=6pt,linestyle=solid,linecolor=gry]{-}(0.8,3)(3.2,3)
\psline[linewidth=6pt,linestyle=solid,linecolor=gry]{-}(1.3,2.5)(2.7,2.5)

\psline[linewidth=0.5pt,linestyle=solid]{-}(0,3)(1,4)
\psline[linewidth=0.5pt,linestyle=solid]{-}(0,2)(2,4)
\psline[linewidth=0.5pt,linestyle=solid]{-}(0,1)(3,4)
\psline[linewidth=0.5pt,linestyle=solid]{-}(0,0)(4,4)
\psline[linewidth=0.5pt,linestyle=solid]{-}(1,0)(4,3)
\psline[linewidth=0.5pt,linestyle=solid]{-}(2,0)(4,2)
\psline[linewidth=0.5pt,linestyle=solid]{-}(3,0)(4,1)

\psline[linewidth=0.5pt,linestyle=solid]{-}(0,1)(1,0)
\psline[linewidth=0.5pt,linestyle=solid]{-}(0,2)(2,0)
\psline[linewidth=0.5pt,linestyle=solid]{-}(0,3)(3,0)
\psline[linewidth=0.5pt,linestyle=solid]{-}(0,4)(4,0)
\psline[linewidth=0.5pt,linestyle=solid]{-}(1,4)(4,1)
\psline[linewidth=0.5pt,linestyle=solid]{-}(2,4)(4,2)
\psline[linewidth=0.5pt,linestyle=solid]{-}(3,4)(4,3)

\multiput(0,0)(0,0.5){9}{\psline[linewidth=0.5pt,linestyle=solid]{-}(0,0)(4,0)}

{\color{white}
\multiput(0,0)(2,0){2}{\circle*{0.15}} \put(3,0){\circle*{0.15}}
\put(2.5,0.5){\circle*{0.15}}
\multiput(1,1)(2,0){2}{\circle*{0.15}}
\multiput(1.5,1.5)(2,0){2}{\circle*{0.15}}
\multiput(0,2)(1,0){2}{\circle*{0.15}}
\multiput(2.5,2.5)(1,0){2}{\circle*{0.15}}
\multiput(1,3)(3,0){2}{\circle*{0.15}}
\put(2.5,3.5){\circle*{0.15}}
\multiput(0,4)(2,0){3}{\circle*{0.15}}
}
\multiput(0,0)(2,0){2}{\circle{0.15}} \put(3,0){\circle{0.15}}
\put(2.5,0.5){\circle{0.15}}
\multiput(1,1)(2,0){2}{\circle{0.15}}
\multiput(1.5,1.5)(2,0){2}{\circle{0.15}}
\multiput(0,2)(1,0){2}{\circle{0.15}}
\multiput(2.5,2.5)(1,0){2}{\circle{0.15}}
\multiput(1,3)(3,0){2}{\circle{0.15}}
\put(2.5,3.5){\circle{0.15}}
\multiput(0,4)(2,0){3}{\circle{0.15}}

\multiput(1,0)(3,0){2}{\circle*{0.15}}
\multiput(0.5,0.5)(1,0){2}{\circle*{0.15}} \put(3.5,0.5){\circle*{0.15}}
\multiput(0,1)(2,0){3}{\circle*{0.15}}
\multiput(0.5,1.5)(2,0){2}{\circle*{0.15}}
\multiput(3,2)(1,0){2}{\circle*{0.15}}
\multiput(0.5,2.5)(1,0){2}{\circle*{0.15}}
\multiput(0,3)(3,0){2}{\circle*{0.15}} \put(2,3){\circle*{0.15}}
\multiput(0.5,3.5)(1,0){2}{\circle*{0.15}} \put(3.5,3.5){\circle*{0.15}}
\multiput(1,4)(2,0){2}{\circle*{0.15}}

\put(2,2){\circle*{0.2}}

\psline[linewidth=0.5pt,linestyle=dotted,linearc=0.45]{-}(0.2,3.8)(3.8,3.8)(3.8,0.2)(3.8,0.2)(0.2,0.2)(0.2,3.8)

{\footnotesize
\put(1.93,2.17){0}
\put(2.26,2.77){\(\mathbf 1\!\!\uparrow_{1}\)}
\put(2.76,3.27){\(\mathbf 2\!\!\uparrow_{1}\)}
\put(3.26,3.77){\(\mathbf 3\!\!\uparrow_{1}\)}
\put(2.26,1.77){\(\mathbf 1\!\!\downarrow_{2}\)}
\put(2.76,1.27){\(\mathbf 2\!\!\downarrow_{2}\)}
\put(3.26,0.77){\(\mathbf 3\!\!\downarrow_{2}\)}
\put(1.26,2.77){\(\mathbf 1\!\!\uparrow_{2}\)}
\put(0.76,3.27){\(\mathbf 2\!\!\uparrow_{2}\)}
\put(0.26,3.77){\(\mathbf 3\!\!\uparrow_{2}\)}
\put(1.26,1.77){\(\mathbf 1\!\!\downarrow_{1}\)}
\put(0.76,1.27){\(\mathbf 2\!\!\downarrow_{1}\)}
\put(0.26,0.77){\(\mathbf 3\!\!\downarrow_{1}\)}

\put(1.45,4.12){\(\mathbf 2\!\uparrow_{1}\!\!+\mathbf 2\!\uparrow_{2}\)}
}

\end{pspicture}
\caption{\label{latticeTexample}
Triangular lattice is built by adding \(\uparrow_{1}\!\!+\!\!\uparrow_{2}\) and \(\downarrow_{1}\!\!+\!\!\downarrow_{2}\)
(left) or by adding \(\uparrow_{1}\!\!+\!\!\downarrow_{2}\) and \(\downarrow_{1}\!\!+\!\!\uparrow_{2}\) (right) to each \(\mathbf v\in\mathbb{Z}^{2}\).
}
\end{figure}
\\
Arc \(\mathcal A_{k}(\mathbf T)\), \(k\) edges away from \(\mathbf v_{0}\!\in\!\mathbf T\), contains the vertices in  \(\mathcal A_{k+i}(\mathbb{Z}^{d})\) traversed via \(\mathcal N_{u}(\mathbf v,\mathbf T)\):
\begin{align}
\mathcal A_{k}(\mathbf T)&\!=\!\!\!\!\!\!\displaystyle\bigcup_{\mathbf v\in\mathcal A_{k-1}(\mathbf T)}\!\!\!\!\!\!\mathcal N_{u}(\mathbf v,\mathbf T):\mathcal A_{1}(\mathbf T)\!=\!\{\uparrow_{1},\uparrow_{2},\uparrow_{1,2}\} \notag \\
n_{k}\!\left(\mathcal A_{k+i}\!\left(\mathbb{Z}^{2}\right)\!,\!\mathbf T\right)&\!=\text{number of paths to }\mathcal A_{k+i}\left(\mathbb{Z}^{2}\right) \notag \\
n_{k}\!\left(\mathcal A_{k}(\mathbf T)\right)&\!=\displaystyle\sum_{i}n_{k}\left(\mathcal A_{k+i}\left(\mathbb{Z}^{2}\right),\mathbf T\right) \notag
\end{align}
For vertices in arc \(\mathcal A_{1}(\mathbf T)\), 2 paths end in \(\mathcal A_{1}(\mathbb{Z}^{2})\) and 1 path ends in \(\mathcal A_{2}(\mathbb{Z}^{2})\):
\begin{align}
\mathcal A_{1}(\mathbf T)&=\{\mathbf v_{0}+\!\!\uparrow_{1},\mathbf v_{0}+\!\!\uparrow_{2},\mathbf v_{0}+\!\!\uparrow_{1,2}\}
\Rightarrow\!\!
\begin{array}{ll}
2&\to_{1} \\
1&\to_{2}
\end{array}
\!\!\Rightarrow\!\!
\begin{array}{ll}
|\!\to_{1}\!|&\!\!=2 \\
|\!\to_{2}\!|&\!\!=1
\end{array}
\notag \\
\to_{1}&=\text{path from }\mathbf v_{0}\text{ to a vertex in embedded }\mathcal A_{1}(\mathbb{Z}^{2}) \notag \\
|\!\to_{1}\!|&=\text{number of paths from }\mathbf v_{0}\text{ to }\mathcal A_{1}(\mathbb{Z}^{2}) \notag
\end{align}
From \(\mathcal A_{1}(\mathbf T)\), the number of paths doubles for up-steps from \(\mathcal A_{1}(\mathbb{Z}^{2})\) to \(\mathcal A_{2}(\mathbb{Z}^{2})\) and from \(\mathcal A_{2}(\mathbb{Z}^{2})\) to \(\mathcal A_{3}(\mathbb{Z}^{2})\).
The number of paths does not change for up-steps from \(\mathcal A_{1}(\mathbb{Z}^{2})\) to \(\mathcal A_{3}(\mathbb{Z}^{2})\) and from \(\mathcal A_{2}(\mathbb{Z}^{2})\) to \(\mathcal A_{4}(\mathbb{Z}^{2})\) :
\begin{align}
\mathcal A_{2}(\mathbf T)&=\!\!\!\!\bigcup_{\mathbf v\in\mathcal A_{1}(\mathbf T)}\!\!\!\!\mathcal N_{u}(\mathbf v,\mathbf T)
\!\Rightarrow\!\!
\begin{array}{ll}
2|\!\to_{1}\!|&.\!\to_{2} \\
1|\!\to_{1}\!|&.\!\to_{3} \\
2|\!\to_{2}\!|&.\!\to_{3} \\
1|\!\to_{2}\!|&.\!\to_{4}
\end{array}
\!\!\Rightarrow\!\!
\begin{array}{l}
|.\!\to_{2}\!|\!=1\!\cdot\!2^{2} \\
|.\!\to_{3}\!|\!=2\!\cdot\!2^{1} \\
|.\!\to_{4}\!|\!=1\!\cdot\!2^{0}
\end{array}
\notag \\
.\!\to_{2}=&\text{path ending in }\mathcal A_{2}(\mathbb{Z}^{2}) \notag \\
|.\!\to_{2}\!|=&\text{number of paths ending in }\mathcal A_{2}(\mathbb{Z}^{2}) \notag
\end{align}
When counting the paths from \(\mathbf v_{0}\) to the arcs \(\mathcal A_{k}(\mathbb{Z}^{2})\), \(\mathcal A_{k+1}(\mathbb{Z}^{2})\), .., \(\mathcal A_{2k}(\mathbb{Z}^{2})\), the coefficients \(c_{i_k}\) follow the rule generated by Pascal's triangle, \footnote{From {\it The On-Line Encyclopedia of Integer Sequences} at http://oeis.org.}: 
\begin{align}
&c_{i_k}\!\!\cdot2^{k-i}\!=\!\text{number of paths to }\mathcal A_{k+i}\left(\mathbb{Z}^{2}\right)\text{ after \(k^{th}\) step} \notag \\
&1^{st}\!\!:\ 1,1 \Rightarrow 1\cdot2^{1}+1\cdot2^{0}\text{ paths to }\mathcal A_{1}(\mathbf T) \notag \\
&2^{nd}\!\!:\ 1,2,1 \Rightarrow 1\cdot2^{2}+2\cdot2^{1}+1\cdot2^{0}\text{ paths to }\mathcal A_{2}(\mathbf T) \notag \\
&3^{rd}\!\!:\ 1,3,3,1 \Rightarrow 1\cdot2^{3}+3\cdot2^{2}+3\cdot2^{1}+1\cdot2^{0}\text{ paths} \notag \\
&4^{th}\!\!:\ 1\!=\!c_{0_4},4\!=\!c_{1_4},6\!=\!c_{2_4},4\!=\!c_{3_4},1\!=\!c_{4_4} \notag \\
&\ \ \ .. \notag \\
&\displaystyle\sum_{i=0}^{i=k}{k\choose i}\!\cdot\!2^{k-i}=\text{number of paths to }\mathcal A_{k}(\mathbf T) \notag \\
&{k\choose i}\!\cdot\!2^{k-i}=\text{number of paths to }\mathcal A_{k+i}\left(\mathbb{Z}^{2}\right) \notag
\end{align}
When odd or even \(k\to\infty\),
\begin{align}
&p_{H}(\mathbf T)\le\text{min \(p\) for which }{k\choose i}\cdot2^{k-i}\cdot p^{k}\ge1 \notag \\
&\displaystyle\binom{k}{\frac{k-1}{2}}\cdot2^{k-\left(\frac{k-1}{2}\right)}=\text{number of paths to }\mathcal A_{k+\frac{k-1}{2}}(\mathbb{Z}^{2}) \notag \\
&\ \ \ \ \ \ \ \ \text{or} \notag \\
&\displaystyle\binom{k}{\frac{k}{2}-1}\cdot2^{k-\left(\frac{k}{2}-1\right)}=\text{number of paths to }\mathcal A_{k+\frac{k}{2}-1}(\mathbb{Z}^{2}) \notag
\end{align}
\begin{align}
\psi(\mathbf T,p)&=
2^{\frac{1}{2}}\lim_{k\to\infty}\left(2^{\frac{1}{2}} \cdot \displaystyle\binom{k}{\frac{k-1}{2}}^{\frac{1}{k}} \cdot p\right)^{k} \notag \\
&\text{or} \notag \\
\psi(\mathbf T,p)&=
2\lim_{k\to\infty}\left(2^{\frac{1}{2}} \cdot \displaystyle\binom{k}{\frac{k}{2}-1}^{\frac{1}{k}} \cdot p\right)^{k} \notag \\
\Rightarrow p_{H}(\mathbf T)&\le\left\{
\begin{array}{l}
\displaystyle\lim_{k\to\infty} \displaystyle\frac{1}{2^{\frac{1}{2}} \cdot \displaystyle\binom{k}{\frac{k-1}{2}}^{\frac{1}{k}}} \\
\text{or} \\
\displaystyle\lim_{k\to\infty} \displaystyle\frac{1}{2^{\frac{1}{2}} \cdot \displaystyle\binom{k}{\frac{k}{2}-1}^{\frac{1}{k}}}
\end{array}
\right.
\approx\displaystyle\frac{1}{2^{\frac{3}{2}}}\approx0.3535 \notag
\end{align}
\(\lim_{k\to\infty}\displaystyle\binom{k}{\frac{k-1}{2}}^{\frac{1}{k}}\approx2\)
, \cite{robbins1955}\footnote{Limits at http://www.wolframalpha.com.}.

\bibliography{bibliography}
\bibliographystyle{unsrt}
\end{document}